\documentclass[twocolumn%
			  ,superscriptaddress%
			  ,aps%
		      ,pra]{revtex4}

\usepackage{amsmath,amssymb}
\usepackage{graphicx}
\begin{document}
	
\title{Quantum revival patterns from classical phase-space trajectories}

\author{Gabriel M. Lando}
\affiliation{Centro Brasileiro de Pesquisas Fisicas,
	Rua Xavier Sigaud 150, 22290-180, Rio de Janeiro, R.J., Brazil}
\author{Ra\'ul O. Vallejos}
\affiliation{Centro Brasileiro de Pesquisas Fisicas,
	Rua Xavier Sigaud 150, 22290-180, Rio de Janeiro, R.J., Brazil}
\author{Gert-Ludwig Ingold}
\affiliation{Universit{\"a}t Augsburg, Institut f{\"u}r Physik, 86135 Augsburg, Germany}
\author{Alfredo M. Ozorio de Almeida}
\affiliation{Centro Brasileiro de Pesquisas Fisicas,
	Rua Xavier Sigaud 150, 22290-180, Rio de Janeiro, R.J., Brazil}
\date{\today}

\begin{abstract}
	A general semiclassical method in phase space based on the final value
	representation of the Wigner function is considered that bypasses
	caustics and the need to root-search for classical trajectories. We
	demonstrate its potential by applying the method to the Kerr
	Hamiltonian, for which the exact quantum evolution is punctuated by a sequence of intricate revival patterns. 
	The structure of such revival patterns, lying far beyond the Ehrenfest time, is semiclassically reproduced and revealed as a consequence of 
	constructive and destructive interferences of classical trajectories.
\end{abstract}

\maketitle  

\section{Introduction}

The semiclassical approximation to quantum mechanics is valuable not only when an exact solution is out of reach, but it also sheds light on a quantum system's classical backbone in both chaotic and integrable scenarios \cite{Gutzwiller}. It was not until the seminal work of Tomsovic and Heller \cite{HelTom}, however, that semiclassical approximations were shown to describe intricate phenomena in non-trivial time regimes. Indeed, such approximations remained valid for longer than the previously established threshold for accuracy, the Ehrenfest time, at which classical structure finer than a quantum cell starts to develop \cite{Gutzwiller, Maia, Roman}. Motivated by this success, semiclassical methods have since been applied for systems with ever-increasing complexity, testing the limit of what one would consider to be exclusively quantum \cite{Zi, Suarez, Sepulveda}. 

Among the class of intrinsically quantum phenomena, the so called quantum revival patterns, also known as fractional revivals \cite{Robinett, Berry}, are a formidable example: They are characterized by shifted and superposed replicas of the initial distribution. Reproducing this phenomenon using the standard semiclassical methods initiated by van Vleck \cite{Van Vleck, Gutzwiller} is problematic mainly because the semiclassical propagator diverges at regions known as caustics, which proliferate in the time interval required for a revival pattern to form. Even though it is possible to use the more sophisticated uniform approximations \cite{Ber76}, which override infinities, they are valid only locally until the next caustic is met. Such approximations are also unable to amend a standard difficulty in semiclassical propagation, known as the root-search problem, which is the need to seek and select relevant classical trajectories based on boundary conditions.

Initial value representations (IVR) \cite{Mil70} have been tailored to deal with such difficulties. The
Herman-Kluk propagator \cite{HerKluk, Kay1, FGrossmann}, for instance, is an integral over
classical trajectories defined by their initial values, requiring no root-search. As an off-shot, the
caustic singularities are replaced by zeros and a
workable approximation for quantum evolution is then achieved \cite{Mil01,Mil12}.

However, in several applications such as the semiclassical treatment of decoherence \cite{Zurek}, it
is desirable to employ methods that have been
developed based on the Weyl representation of quantum mechanics. Here, one
evolves directly the Wigner function \cite{Wigner, Rios, Dittrich} or its Fourier transform,
the chord function, by the Weyl transform of the semiclassical propagator
\cite{Ber89, Report, Brodier}. A mere change of variables to initial or final values of
the trajectories results in divergence-free phase space evolution \cite{IVRFVR}.
Thus, the desirable features of previous IVR techniques follow naturally, without any need
to substitute the standard semiclassical propagators that are derived directly from 
path integrals \cite{Gutzwiller}.

Whether or not an initial (or final) value representation will be able to remain accurate and unshaken by caustics,
which usually spoil standard propagation, must be checked numerically.
Since the time evolution under the action of quadratic Hamiltonians is semiclassically exact, such systems are not suited as a testbed. The difficulty is then that higher-order Hamiltonians,
which generate non-linear classical motion, do not generally have exactly solvable quantum
evolutions. How can one then be certain about features in the semiclassical evolution 
without an exact result with which to compare them? Indeed, one finds that a vast literature
has grown that is based on the convenience of these methods without ever addressing 
this basic dilemma.

Here, we propose the square of the simple harmonic oscillator, the 4th-order Kerr system,
as the ideal benchmark test for a semiclassical method for systems with one degree of freedom. 
It is singled out by the following properties: (a) One
can calculate exactly its intricate quantum evolution, which displays a
periodic structure of the aforementioned quantum revival patterns; (b) classical
trajectories can be obtained analytically, thus avoiding integration errors;
(c) the caustic structure is so complex that it must reveal any shortcoming in
the method. Also, since semiclassical methods are only exact for quadratic Hamiltonians, 
the semiclassical treatment of Kerr propagation is still only an approximation,
regardless of the exactness of its classical trajectories.
This is the stringent test to which we submit the recently proposed phase space 
final value representation (FVR) \cite{IVRFVR}.

The Kerr system is also of significant experimental interest. The optical
Kerr effect can be emulated, e.g., in a Bose-Einstein condensate confined by a
three-dimensional optical lattice \cite{greiner02}. A three-dimensional circuit
quantum electrodynamic architecture was also used to engineer an artificial
Kerr medium in order to observe fractional revivals of a coherent state
\cite{kirchmair13}. Due to the extremely weak nonlinearities of most materials,
however, collapses and revivals due to the Kerr effect have not yet been
observed in optical media.

Direct use of the semiclassical propagator by Toscano \textit{et al.} for the
Kerr system did allow for accurate evaluation of the autocorrelation function
of a coherent state for long times \cite{toscano09}, but the wave functions
were only evaluated far from caustics. Tomsovic \textit{et al.} used
sophisticated complex time-dependent semiclassical propagators to accurately
calculate autocorrelation functions of Gaussian many-body states of
Bose-Hubbard systems with their Kerr-like Hamiltonian beyond the Ehrenfest time
\cite{tomsovic18}. The many-body context was also the motivation of the cruder
semiclassical approximation for the Wigner function propagator in
\cite{mathew18}, which nonetheless did detect full revivals for the Kerr
system through the annihilation operator's expectation value for an initial
coherent state. Moreover, the Herman-Kluk approximation was applied to the
$0$-dimensional Bose-Hubbard model: By lifting the wave function into 
phase space, simple interferences were visually captured for an initial
coherent state placed very close to the origin in a Kerr-like system \cite{grossman}.
However, none of the previous explorations attain the high degree of detailed verification
of a semiclassical method as we here exhibit for the Kerr evolution.

The text is organized as follows: In Sec. II we introduce the Kerr system, presenting its exact classical and quantum evolutions. This is followed by a description of the FVR method in Sec. III and its results in Sec. IV. We discuss the semiclassical mechanism for revival production in Sec. V and finish the paper with the final remarks of Sec. VI. Movies of quantum and semiclassical evolution of the Wigner function for a coherent state are provided as Supplemental Material.

\section{The Kerr System}

The 4th-order Kerr Hamiltonian which we consider here is essentially the square
of the Hamiltonian of a simple harmonic oscillator. With an appropriate choice of units
for position, momentum, and energy, we can always bring the Kerr Hamiltonian into
the form
\begin{equation}
H(q, p) = (p^2+q^2)^2\,.
\label{eq:HKerr}
\end{equation}

\begin{figure}[h!]
	\includegraphics[width=\linewidth]{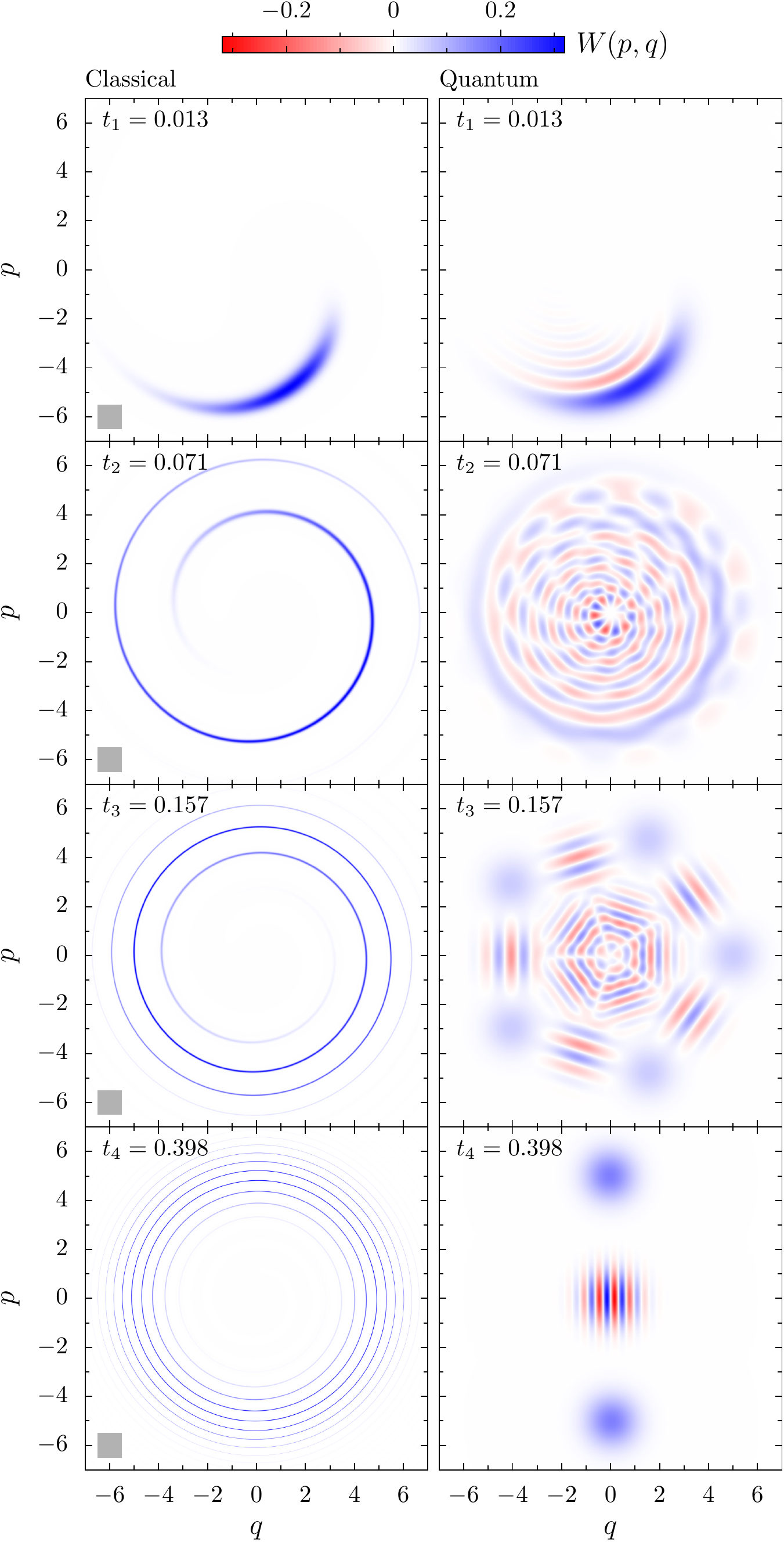}
	\caption{The classical and quantum Wigner functions of a coherent state initially centered at the phase-space
		point $(q, p) = (5, 0)$ are displayed for four different times, with the largest one exceeding $ 6 T_E$, the Ehrenfest time for this case being given by \eqref{eq:ehren} as $T_E \approx 0.063$. The times $t_3 = T_\text{rev}/5$ and $t_4 = T_\text{rev}/2$ correspond to fractional revivals, where $T_\text{rev}$ is given by \eqref{eq:rev}. The gray squares in the classical plots represent regions of area $\hbar$.}
	\label{fig:wigner_cl_qu_sc}
\end{figure}

Viewing (\ref{eq:HKerr}) as a classical Hamiltonian, one finds that 
\begin{equation}
\omega = 4 (p^2+q^2)  
\end{equation}
is conserved for each orbit, playing the role of an angular frequency. The resulting Hamilton equations of motion can then be solved analytically. Since orbits with a larger radius have higher angular velocities, the initial classical distribution will both revolve around the origin and stretch into a thin filament, as can be seen in the evolution of a coherent state displayed in the left column of Fig.~\ref{fig:wigner_cl_qu_sc} for four different times $t$.

Quantum mechanically, $q$ and $p$ in (\ref{eq:HKerr})
become operators satisfying the commutation relation $[\hat q,\hat p] = i$, where we adopt $\hbar=1$.
Introducing the number operator $\hat n$ of the harmonic oscillator, we can
express the Kerr Hamiltonian as
\begin{equation}
H = (2\hat n+1)^2
\label{eq:HKerr_n}
\end{equation}
with its Fock eigenstates $\vert n\rangle$.

It has long been noticed \cite{Stoler1} that under the action of the Kerr
Hamiltonian (\ref{eq:HKerr_n}) fractional revivals occur in the quantum evolution
of a coherent state
\begin{equation}
\vert \alpha \rangle = e^{-\frac{\vert \alpha \vert^2}{2}}
\sum_{n=0}^\infty \frac{\alpha^n}{\sqrt{n!}} \vert n \rangle \,
\label{coher}
\end{equation}
where $\alpha = \left( \langle\hat{q}\rangle + i \langle\hat{p}\rangle
\right)/\sqrt{2}$. Following \cite{Aver1}, we choose times $t=(2a/b)
T_\text{rev}$, where the integers $a$ and $b$ are mutually prime and the time for a complete revival is given by 
\begin{equation}
T_\text{rev}= \frac{\pi}{4} \, . \label{eq:rev}
\end{equation}
By choosing sufficiently large integers, any given time can be approximated to the desired
accuracy. It can be shown \cite{Aver1} that for times $t$ of the form introduced
here, the evolved coherent state $\vert\alpha(t)\rangle$ can be expressed as a
superposition of $b$ or $b/2$ coherent states when $b$ is odd or even,
respectively. In particular, for $a=1$ and $b=2m$, one finds a fractional revival
pattern of order $m$. 

The classical evolution of the Wigner function for a coherent state \eqref{coher} is obtained by solving the Liouville equation or, equivalently, by propagating the initial Wigner function using the classical equations of motion. The quantum evolution is given by time-evolving \eqref{coher} under \eqref{eq:HKerr_n} and taking its well-known Wigner transform \eqref{wig}, to be	 introduced in the next section. In Fig.~\ref{fig:wigner_cl_qu_sc} we provide the classical (left column) and quantum (right column) exact Wigner functions for the evolution of an initial coherent state at four distinct times.  We see that for $t_1$ the classical backbone is clearly visible in the quantum Wigner function, together with the typical interference patterns. Such clear analogies between classical and quantum behavior are expected up to the Ehrenfest time,
\begin{equation}
T_E = \frac{2 \pi}{\omega_c}, \label{eq:ehren}
\end{equation}
in which the initial distribution's center, moving with angular velocity $\omega_c$, has performed a full revolution around the origin~\cite{Roman}. For $t_2$, for instance, we have already exceeded $T_E$ and the superposition of multiple interferences masks the relationship to the underlying classical structure. The following panels for $t_3$ and $t_4$ display fractional revival patterns. Their time values of $t_3 = T_\text{rev}/5$ and $t_4 = T_\text{rev}/2$ by far surpass the Ehrenfest time: Not only has the classical filament become quite thin, but the gap between different windings has narrowed to $\mathcal{O}(\hbar^{1/2})$.

\section{Evolution of the Wigner Function}

Our focus is the analysis of the intricate
features of the evolved state $|\psi(t)\rangle$,
best examined within a full representation in phase space with coordinates
$(q,p)$, the position and the momentum, respectively.
Among these, the Weyl representation, whose main object is the Wigner function
\begin{equation}
W (q, p, t) = \frac{1}{\pi \hbar} \! \int \! d\tilde q\langle q+\tilde q | \psi(t) \rangle\langle\psi (t) | q-\tilde q\rangle
e^{-2i\tilde q p/ \hbar}\,, \label{wig}
\end{equation}
is equivalent to, e.g, the position or momentum representations of quantum mechanics.
In particular, position marginal distributions are given by
\begin{equation}
\vert \langle q|\psi(t)\rangle \vert^2 = \int dp \, W (q,p,t) \, ,
\label{eq:marginal}
\end{equation}
and analogously for momentum. The squared autocorrelation is also obtainable from the Wigner function as
\begin{equation}
A^2 (t) = 2 \pi \int dq dp \,W(q,p,t) W (q,p,0) \, .
\label{eq:autoc}
\end{equation}
Hence, the portrait of the evolved Wigner function is sufficient for testing
any semiclassical method. 

The divergences of standard semiclassical methods can be traced to the 
amplitude in the propagator as caustics are approached.
A change of variables in the several alternative methods proposed in \cite{IVRFVR} suppress these divergences, with the further advantage that the FVR preserves its semiclassical form for purely quadratic Hamiltonians. 

\begin{figure}
	\includegraphics[width=.7\linewidth]{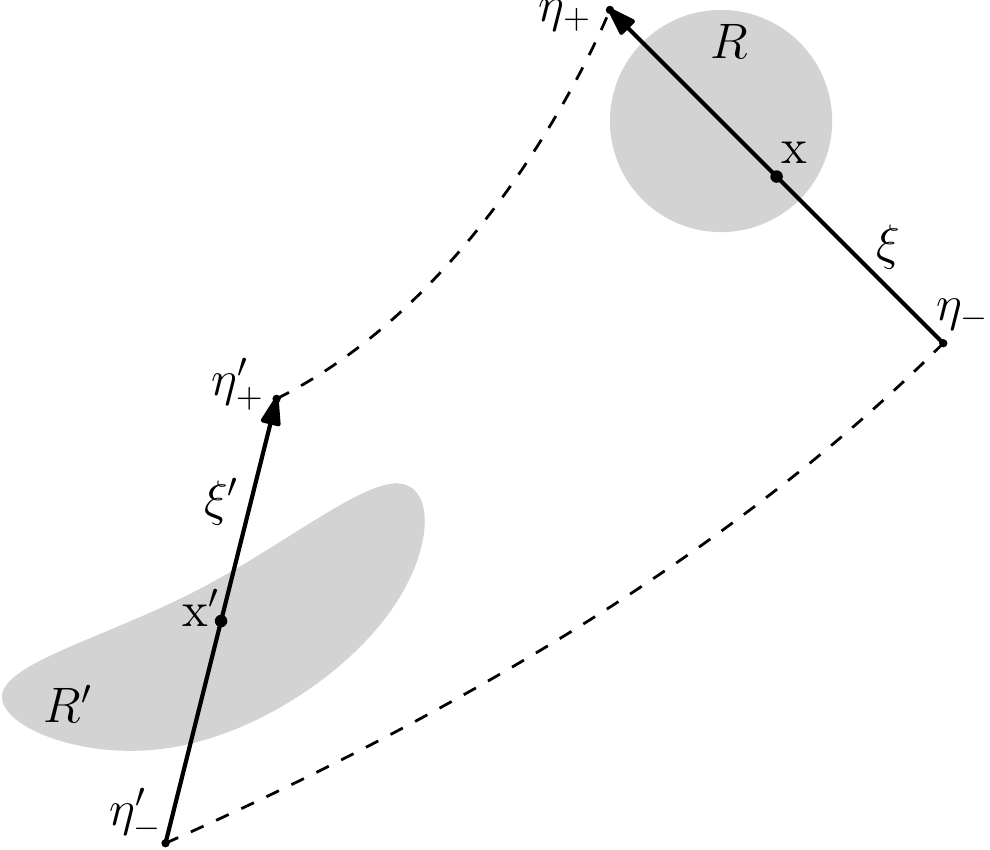}
	\caption{A final chord $\xi'$, with center $\text{x}'$, is evolved backwards
		to form an initial chord $\xi(\xi', \text{x}', t)$, centered at
		$\text{x}(\xi',\text{x}',t)$. By considering the endpoints of $\xi'$ and $\xi$,
		such evolution is described by the circuit 
		$\eta_-' \mapsto \eta_- \mapsto \eta_+ \mapsto \eta_+'$. The
		first and last propagations are performed along classical trajectories (dashed lines), 
		while the middle one is a reflection around $\text{x}$, given by $\eta_+ = 2\text{x} - \eta_-$. The shaded areas $R$ and $R'$ represent the Wigner function at initial and final times, respectively.}
	\label{fig:quad}
\end{figure}

The time evolved Wigner function within the FVR approach depends on pairs of trajectories with 
initial values $\eta_\pm$ and final values $\eta_\pm'$, as depicted in Fig.~\ref{fig:quad}. The initial center and chord 
are given by $\text{x} = (\eta_+ + \eta_-)/2$ and $\xi = \eta_+ - \eta_-$, respectively; the final center and chord, $\text{x}'$ and $\xi'$, are defined accordingly. Then, following \cite{IVRFVR},
\begin{equation}
W(\text{x}',t) =\!\! \int \! \frac{d\xi_p'd\xi_q'}{(2\pi)} \left\vert \det \frac{d \xi }{d \xi'} \! \right\vert^\frac{1}{2} \! \! \! \exp \left\{ \! i \! \left[ \tilde{S}_{\text{x}'}(\xi) \! + \!  \frac{\tilde{\sigma} \pi}{2}\right] \! \right\} \! \chi \left( \xi \right) \, , \label{fvr}
\end{equation}
\begin{figure}
	\includegraphics[width=\linewidth]{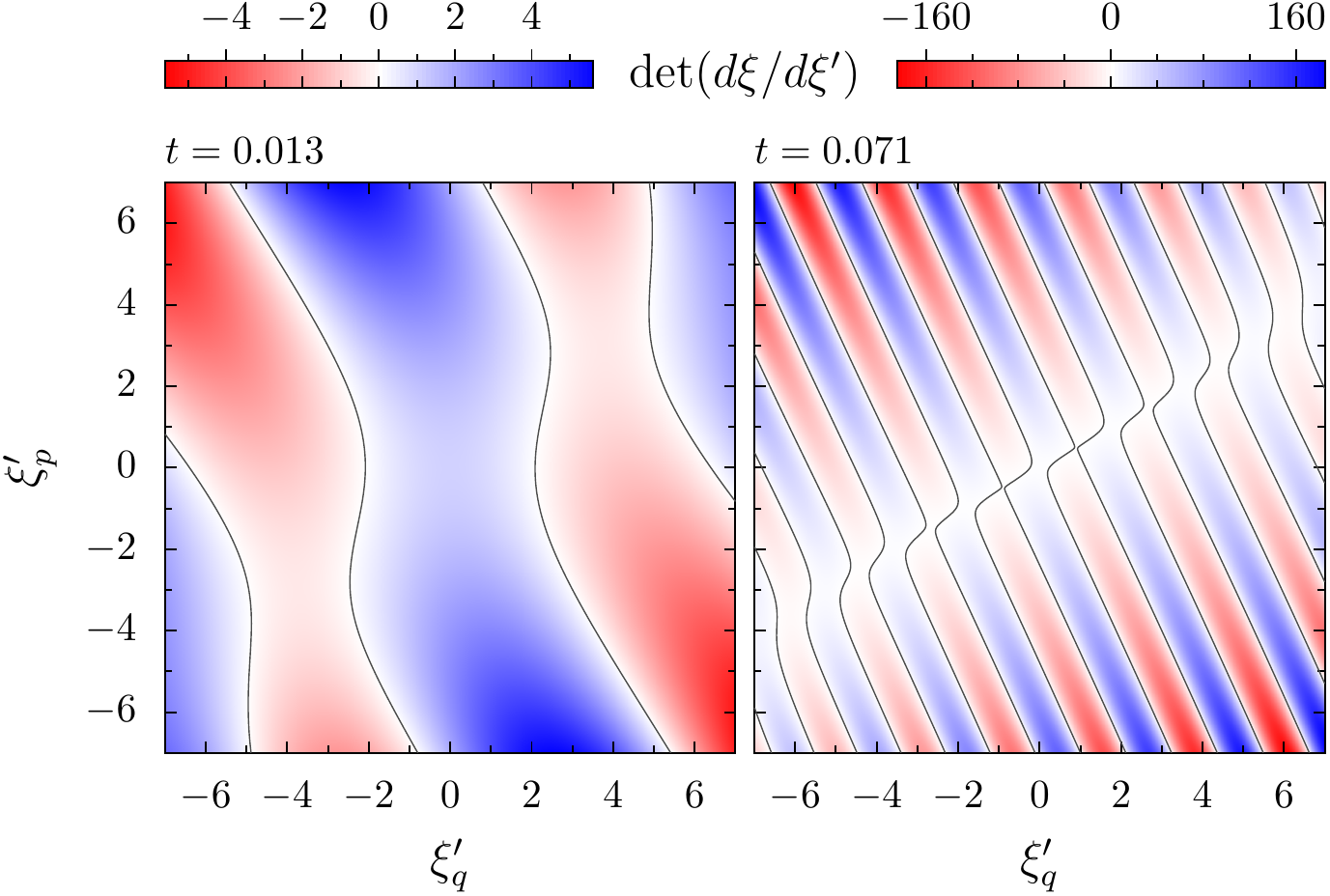}
	\caption{Density plot for the determinant in \eqref{fvr} in the
		$(\xi_q',\xi_p')$-plane at times $t=0.013$ and $t=0.071$ for the final
		Wigner function evaluated at $(q,p)=(5,2)$. The caustic
		submanifolds at which the original root-search based propagator diverges are displayed as solid black curves. The initial state is
		the same as for Fig.~\ref{fig:wigner_cl_qu_sc}.}
	\label{fig:caustics}
\end{figure}
$\!\!$which is expressed in terms of functions that we now describe. The argument $\xi$ of both $\chi$ and the action $ \tilde{S}_{\text{x}'}$ is defined in the present FVR as the backward classical propagation of the integration variables $\xi' = (\xi'_q, \xi'_p)$. The initial state, on which absolutely no restriction is placed, enters through the symplectic Fourier transform of its Wigner function at $t=0$,
\begin{equation}
\chi \left( \xi \right) = \int \frac{dy}{(2\pi)} \, \exp \left( -i y \cdot J \xi \right) W(y,0)\,,
\end{equation}
$J$ being the symplectic matrix. Note that, for coherent states, both $\chi(\xi)$ and $W(y,0)$ are Gaussians. The exponential in (\ref{fvr}) arises from a
semiclassical approximation of an evolved phase space reflection \cite{IVRFVR,
	Brodier}. The choice of final instead of initial values as integration variables is
just a matter of choice when classical trajectories are exact, but preferable
when only numerical solutions are available \cite{footnote2}. Considering 
that the initial chord $\xi$ is determined by the pair of trajectories 
propagated backward from $\eta_\pm'$,
the first term in the phase of (\ref{fvr}) can be written as
\begin{equation}
\tilde{S}_{\text{x}'}(\eta'_\pm, \eta_\pm) = \eta_+ \cdot J \eta_- + t \Delta H(\eta_\pm) - \oint_{\mathcal{C}(\eta_\pm',\eta_\pm)} p \, dq    \, . \label{sss}
\end{equation}
Here, $\Delta H(\eta_\pm) = H(\eta_+) - H(\eta_-)$ is the energy difference between the endpoints of $\xi$ and $\mathcal{C} (\eta_\pm',\eta_\pm) $ is the closed contour given by $\eta_-' \mapsto \eta_- \mapsto \eta_+ \mapsto \eta_+' \mapsto \eta_-'$, detailed in Fig.~\ref{fig:quad}.	  
The term $\tilde{\sigma}$ in (\ref{fvr}) counts the zeros of the
determinant in the prefactor up to time $t$ and is related to the Maslov index
\cite{Littlejohn, OAI}. The regions defined by this determinant's kernel are
exactly the caustic submanifolds, which form a complex web in the Kerr system,
as can be seen in Fig.~\ref{fig:caustics}. 

\section{Semiclassical Evolution for the Kerr Hamiltonian}

\begin{figure}
	\includegraphics[width=\linewidth]{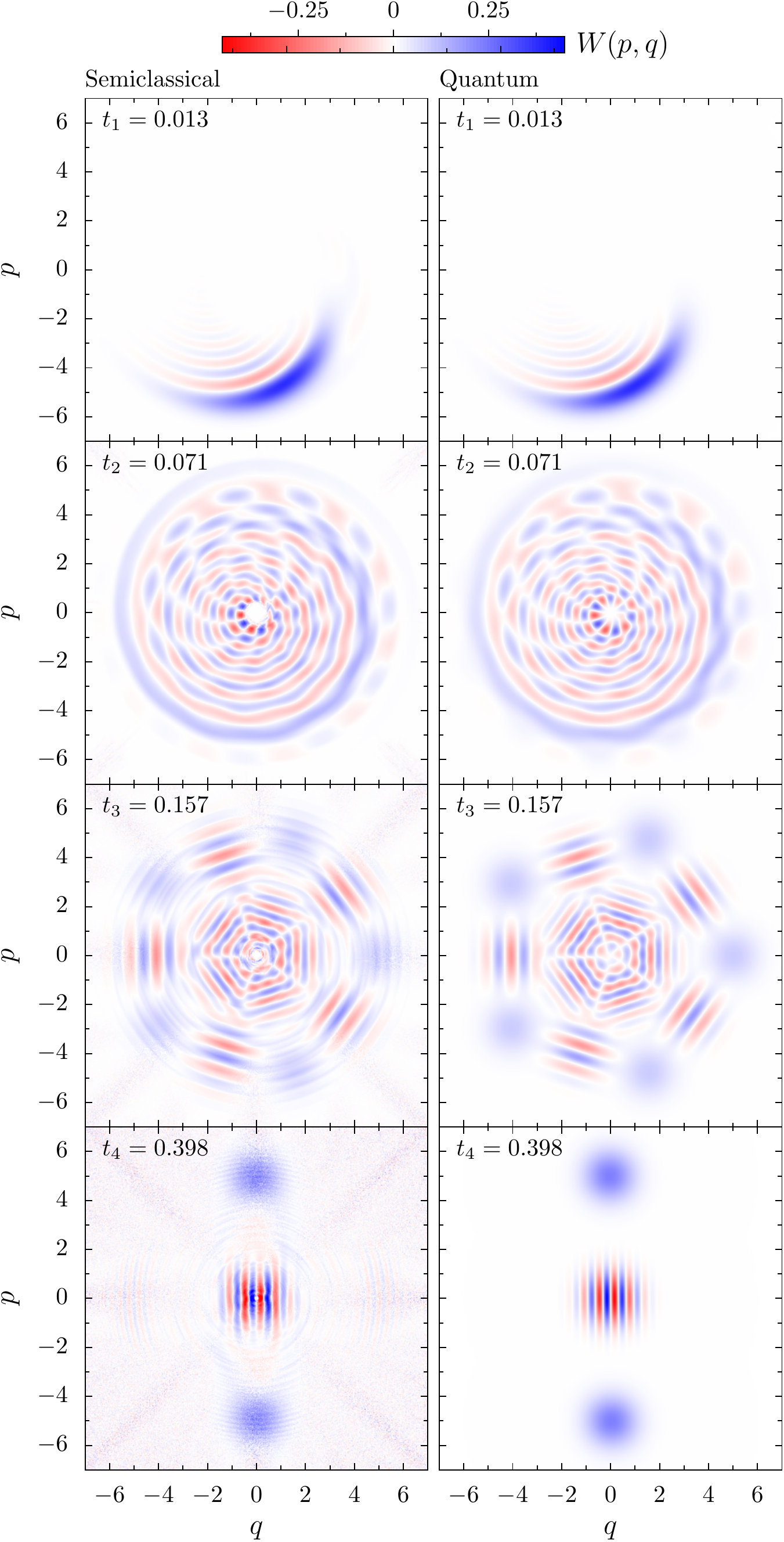}
	\caption{The semiclassical Wigner functions obtained from the FVR in Eq.~\eqref{fvr} (left column) for the same time values as in Fig.~\ref{fig:wigner_cl_qu_sc}. We repeat the exact quantum equivalents for comparison (right column). Notice how the FVR is able to transform the thin classical spirals in Fig.~\ref{fig:wigner_cl_qu_sc} into pentagonal and cat states. For more details about the time evolution see the movies in the Supplemental Material.}
	\label{fig:wigner_sc}
\end{figure}

Semiclassical propagation
depends entirely on classical trajectories and must not only reproduce the
interferences between the classical whirl and itself, but also eventually
cancel them in large phase-space regions at fractional revival patterns (cf. panels for $t_3$ and $t_4$ in the right column of Fig.~\ref{fig:wigner_cl_qu_sc}). Fig.~\ref{fig:wigner_sc} shows that, despite the complex caustic
structure and large interference-free regions, the FVR method successfully
reproduces fractional revival patterns that occur for times much longer than
the Ehrenfest time. 

\begin{figure}
	\includegraphics[width=\linewidth]{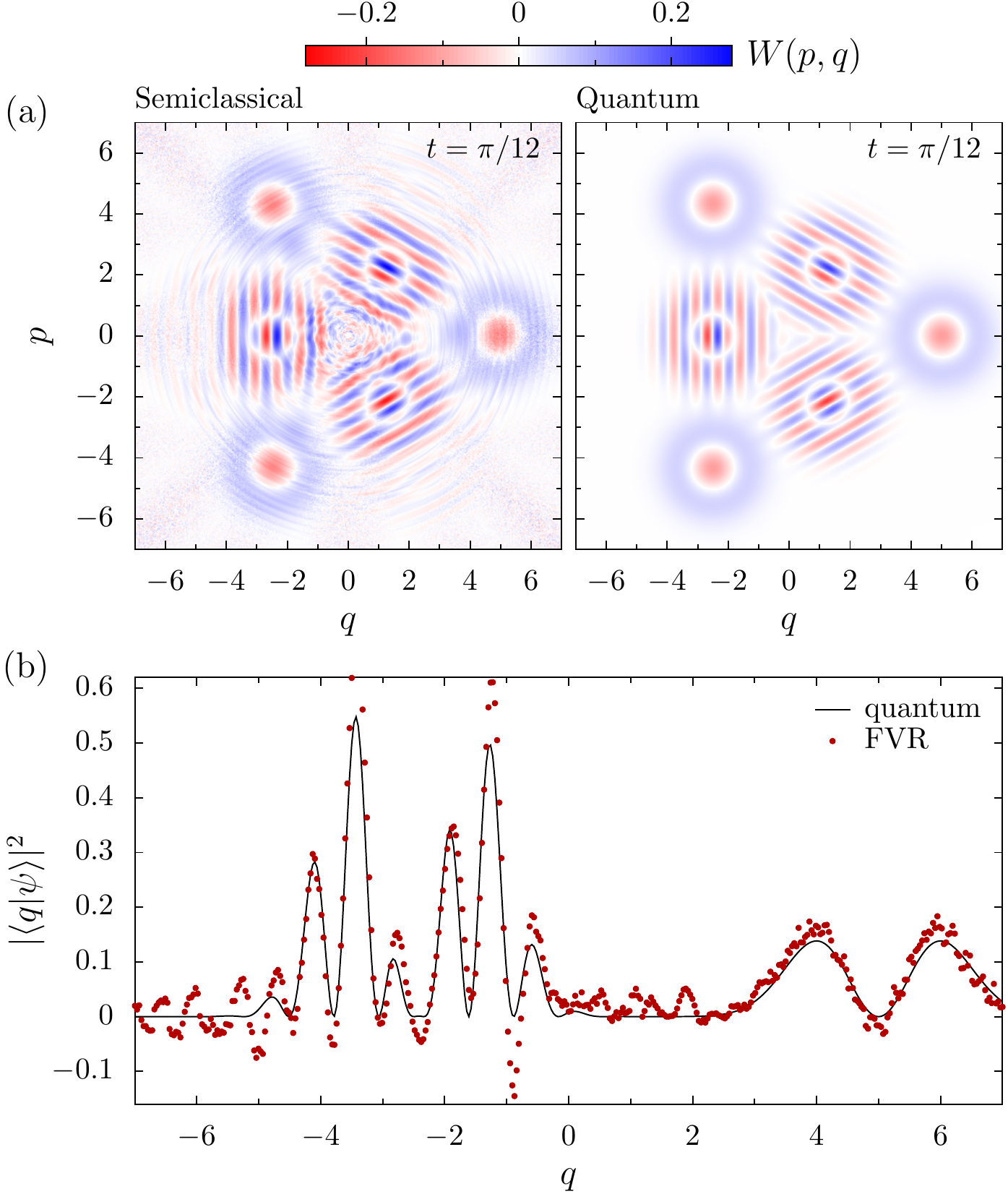}
	\caption{(a) Semiclassical (left) and quantum (right) Wigner functions for the triangular revival at $t=\pi/12$. The
		initial state is a displaced $n=1$ Fock state centered at $(q,p)=(5,0)$. (b) Position marginal distributions obtained from \eqref{eq:marginal} for the post-normalized Wigner functions displayed in (a).}
	\label{fig:fock}
\end{figure}

It should be recalled that standard semiclassical methods based on root-search
are limited to initial states that are either initial coherent states \cite{HelTom, Littlejohn} or approximate WKB
states \cite{Gutzwiller, Maia}. A shifted first excited Fock state lies	
outside both these classes, but as Fig.~\ref{fig:fock} demonstrates, our FVR
approximation captures its full time evolution just as easily as that of initial
coherent states.

\begin{figure}
	\includegraphics[width=\linewidth]{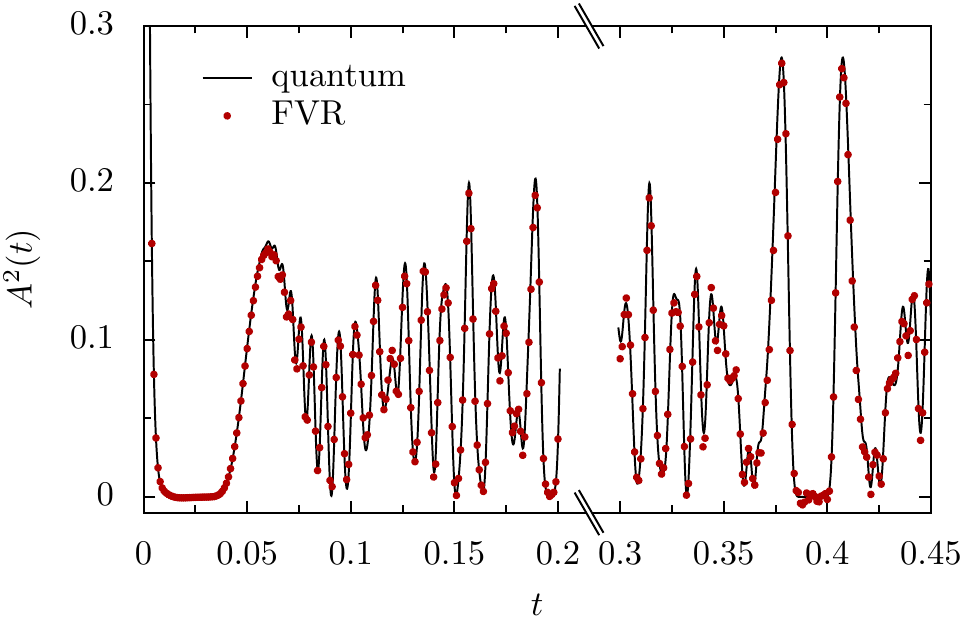}
	\caption{Comparison of the time dependence of the squared autocorrelation function
		\eqref{eq:autoc} for the quantum result (line) and the semiclassical result 
		after post-normalization (points) for the initial state used in Fig.~\ref{fig:wigner_cl_qu_sc}.
		Beyond half the revival time $t=\pi/8 \approx 0.393$, $A^2 (t)$ is seen to
		continue symmetrically. In order to make the fine structure more visible, we have left out part of the time interval.}
	\label{fig:autocorr}
\end{figure}

The squared autocorrelation obtained from the semiclassical and quantum Wigner functions using \eqref{eq:autoc} is displayed in Fig.~\ref{fig:autocorr}. It again affirms the FVR accuracy. Note that due to the fine semiclassical undulations inside the revived coherent states (see Fig.~\ref{fig:wigner_sc}), the Wigner function loses a small fraction of its normalization -- a homogeneous loss that can be corrected by simple post-normalization. This loss is due to a subset of trajectories that do not contribute to the mechanism described in the following section.

\section{Semiclassical Mechanism for Revival Patterns}

Some light can be shed on how the classical trajectories combine to create revivial patterns. We start with the full revival at $T_\text{rev} = \pi/4$. Here, the final Wigner function is equal to the initial one, implying that the final chords $\xi'$ must be backwards evolved near themselves, i.e. $\xi' \approx \xi$. Since the orbits are circles, this condition  fixes contributing $\xi'$ chords as the ones whose endpoints $\eta_+'$ and $\eta_-'$ perform an integer number of complete revolutions around the origin, finishing near their initial values $\eta_+$ and $\eta_-$. In terms of the orbits' angular frequencies,
\begin{align}
\omega_\pm T_\text{rev} = 4|\eta_\pm|^2 T_\text{rev}  = 2 \pi j_\pm \, , \label{eq:times}
\end{align}
where we define the winding numbers $j_\pm$ for $\eta_\pm$. The pairs of orbits whose radii lie in between the successive ``time-quantized" values \eqref{eq:times} define long chords with rapidly oscillating phases that cancel out: Their absence is responsible for the fine undulations in the revived coherent states and affects the Wigner function's normalization.  

Substituting the variables in \eqref{eq:times} in Eq.~\eqref{sss},
\begin{equation}
\tilde{S}_{\text{x}'}(\eta'_\pm, \eta_\pm) - \eta_+ \cdot J \eta_- =  \left( \frac{\pi^2}{4 T_\text{rev}} \right)(j_+^2 - j_-^2) \, ,
\end{equation}
and since the winding numbers for this case are exactly the Maslov contributions, the final phase [modulo a symplectic Fourier transform] is finally given by
\begin{equation}
\tilde{S}_{\text{x}'}(\eta'_\pm, \eta_\pm) + \frac{\tilde{\sigma} \pi}{2} = \pi (j_+-j_-) [1 + (j_+ + j_-)] \, . \label{eq:phase}
\end{equation}
Since $j_+ - j_-$ and $j_+ + j_-$ have the same parity, the phase in \eqref{eq:phase} is always an even multiple of $\pi$. The relevant final chords are, therefore, selected such that the final Wigner function is localized exactly over the initial one, reproducing the complete revival as expected. 

For fractional revivals with times $t = \pi/\beta$, the only difference is that the relevant final chords might perform fractional revolutions around the origin. We can express this condition as
\begin{align}
\frac{\omega_\pm \pi}{\beta}  = \frac{4|\eta_\pm|^2 \pi}{\beta}  = 2 \pi (j_\pm + \alpha) \, , \label{eq:times2}
\end{align}
where $\alpha$ is a rational number that reflects the positions of the final coherent states and thus depends on $\beta$. In contrast to the revived coherent states, the interference patterns appearing near the origin for the cat state revival ($\beta = 8$) are due exclusively to long chords, typically spanning the diameter of the classical spiral.

\section{Conclusion}

The ability to reproduce the complete kaleidoscopic evolution for the quantum Kerr system on the basis of classical phase space trajectories is indisputable evidence for our FVR method. The undulations visible in the semiclassical Wigner functions play no role in extracting typical quantum mechanical quantities obtained by integration, a process which filters out residual classical fine structures. Preliminary results evince that our FVR method is not restricted to exact classical systems, demonstrating the generality of this approach \cite{new}.

\section*{Acknowledgements}

We thank Jonas Bucher and Fabricio Toscano for stimulating discussions. Partial financial
support from CNPq and the National Institute for Science and Technology:
Quantum Information is gratefully acknowledged.


\begin{thebibliography}{99}
	
	\bibitem{Gutzwiller}
	M. Gutzwiller,
	\textit{Chaos in Classical and Quantum Mechanics} (Springer, New York, 1990).
	
	\bibitem{HelTom}
	S. Tomsovic and E. J. Heller,
	Semiclassical Dynamics of Chaotic Motion: Unexpected Long-Time Accuracy,
	Phys. Rev. Lett. \textbf{67}, 664 (1991).
	
	\bibitem{Maia}
	R. N. P. Maia, F. Nicacio, R. O. Vallejos, and F. Toscano,
	Semiclassical Propagation of Gaussian Wave Packets,
	Phys. Rev. Lett. {\bf 100}, 184102 (2008).
	
	\bibitem{Roman}
	R. Schubert, R. O. Vallejos, and F. Toscano,
	How Do Wave Packets Spread? Time Evolution on Ehrenfest Time Scales,
	J. Phys. A {\bf 45}, 215307 (2012).
	
	\bibitem{Zi}
	Z.-M. Lu and M. E. Kellman,
	Phase space structure of triatomic molecules,
	J. Chem. Phys. {\bf 107}, 1 (1997).
	
	\bibitem{Suarez}
	I. M. Suarez Barnes, M. Nauenberg, M. Nockleby, and S. Tomsovic,
	Semiclassical Theory of Quantum Propagation: The Coulomb Potential,
	Phys. Rev. Lett. {\bf 71}, 1961 (1993). 
	
	\bibitem{Sepulveda}
	M. A. Sep\'ulveda and E. J. Heller,
	Semiclassical calculation and analysis of dynamical systems with mixed phase space,
	J. of Chem. Phys. {\bf 101}, 8004 (1998).
	
	\bibitem{Robinett}
	R. W. Robinett,
	Quantum wave packet revivals,
	Phys. Rep. \textbf{392}, 1 (2004).
	
	\bibitem{Berry}
	M. V. Berry, I. Marzoli, W. Schleich, 
	Quantum carpets, carpets of light, 
	Phys. World \textbf{14}(6), 39 (2001).
	
	\bibitem{Van Vleck}
	J. H. van Vleck,
	The Correspondence Principle in the Statistical Interpretation of Quantum Mechanics,
	Proc. Natl. Acad. Sci. USA \textbf{14}, 178 (1928).
	
	\bibitem{Ber76}
	M. V. Berry,
	Waves and Thom's theorem,
	Adv. Phys. \textbf{25}, 1 (1976).
	
	\bibitem{Mil70}
	W. H. Miller,
	Classical \textit{S} Matrix: Numerical Application to Inelastic Collisions,
	J. Chem. Phys. \textbf{53}, 3578 (1970).
	
	\bibitem{HerKluk}
	M. F. Herman and E. Kluk,
	A semiclassical justification for the use of non-spreading wavepackets in dynamics calculations,
	Chem. Phys. \textbf{91}, 27 (1984).
	
	\bibitem{Kay1}
	K. G. Kay,
	Integral expressions for the semiclassical time-‐dependent propagator,
	J. Chem. Phys. \textbf{100}, 4377 (1994).
	
	\bibitem{FGrossmann}
	F. Grossmann,
	Semiclassical coherent-state path integrals for scattering,
	Phys. Rev. A \textbf{57}, 3256 (1998).
	
	\bibitem{Mil01}
	W. H. Miller,
	The semiclassical initial value representation: a potentially practical
	way for adding quantum effects to classical molecular dynamics simulations,
	J. Phys. Chem. A \textbf{105}, 2942 (2001).
	
	\bibitem{Mil12}
	W. H. Miller,
	Perspective: Quantum or classical coherence?,
	J. Chem. Phys. \textbf{136}, 210901 (2012).
	
	\bibitem{Zurek}
	W. H. Zurek,
	Decoherence, einselection, and the quantum origins of the classical,
	Rev. Mod. Phys. \textbf{75}, 715 (2003).
	
	\bibitem{Wigner}
	E. Wigner,
	On the Quantum Correction For Thermodynamic Equilibrium,
	Phys. Rev. \textbf{40}, 749 (1932).
	
	\bibitem{Rios}
	P. P. de M. Rios, and A. M. Ozorio de Almeida,
	On the propagation of semiclassical Wigner functions,
	J. Phys. A: Math. Gen. \textbf{35}, 2609 (2002).
	
	\bibitem{Dittrich}
	T. Dittrich, C. Viviescas, and L. Sandoval,
	Semiclassical Propagator of the Wigner Function,
	Phys. Rev. Lett. \textbf{96}, 070403 (2006).
	
	\bibitem{Ber89}
	M. V. Berry,
	Quantum Scars of Classical Closed Orbits in Phase Space,
	Proc. R. Soc. Lond. A \textbf{423}, 219 (1989).
	
	\bibitem{Report}
	A. M. Ozorio de Almeida,
	The Weyl representation in classical and quantum mechanics,
	Phys. Rep. \textbf{295}, 265 (1998).
	
	\bibitem{Brodier}
	A. M. Ozorio de Almeida and O. Brodier,
	Phase space propagators for quantum operators,
	Ann. Phys. \textbf{321}, 1790 (2006).	
	
	\bibitem{IVRFVR}
	A. M. Ozorio de Almeida, R. O. Vallejos, and E. Zambrano,
	Initial or final values for semiclassical evolutions in the Weyl-Wigner representation,
	J. Phys. A: Math. Theor. \textbf{46}, 135304 (2013).
	
	\bibitem{Stoler1}
	B. Yurke and D. Stoler,
	Generating Quantum Mechanical Superpositions of Macroscopically
	Distinguishable States via Amplitude Dispersion,
	Phys. Rev. Lett. \textbf{57}, 13 (1986).
	
	\bibitem{Aver1}
	I. S. Averbukh and N. F. Perelman,
	Fractional Revivals: Universality in the Long-Term Evolution of
	Quantum Wave Packets Beyond the Correspondence Principle Dynamics,
	Phys. Lett. A \textbf{139}, 449 (1989).
	
	\bibitem{greiner02} 
	M. Greiner, O. Mandel, T. W. H\"ansch, and I. Bloch,
	Collapse and revival of the matter wave field of a Bose-Einstein condensate,
	Nature {\bf 419}, 51 (2002).
	
	\bibitem{kirchmair13}
	G. Kirchmair, B. Vlastakis, Z. Leghtas, S. E. Nigg, H. Paik, E. Ginossar, M. Mirrahimi, L. Frunzio, S. M. Girvin, and R. J. Schoelkopf,
	Observation of quantum state collapse and revival due to the single-photon Kerr effect,
	Nature {\bf 495}, 205 (2013).
	
	\bibitem{toscano09}
	F. Toscano, R. O. Vallejos, D. Wisniacki,
	Semiclassical description of wave packet revival,
	Phys. Rev. E. {\bf 80}, 046218 (2009).
	
	\bibitem{tomsovic18}
	S. Tomsovic, P. Schlagheck, D. Ullmo, J.-D. Urbina, and K. Richter,
	Post-Ehrenfest many-body quantum interferences in ultracold atoms far out of equilibrium,
	Phys. Rev. A \textbf{97}, 061606(R) (2018).
	
	\bibitem{mathew18}
	R. Mathew and E. Tiesinga,
	A semiclassical theory of phase-space dynamics of interacting bosons,
	arXiv:1803.05122v1.
	
	\bibitem{grossman}
	S. Ray, P. Ostmann, L. Simon, F. Grossmann, and W. T. Strunz,
	Dynamics of interacting bosons using the Herman-Kluk semiclassical initial value representation,
	J. Phys. A: Math. Theor. \textbf{49}, 165303 (2016).
	
	\bibitem{footnote2}
	By considering final instead of initial values we can backwards propagate both tips of $\xi' \mapsto \xi$ only once, instead of propagating the initial value $\eta_-$ forwards and then backwards along the cycle $\eta_- \mapsto \eta_-' \mapsto \eta_+' \mapsto \eta_+$, which increases propagation errors.
	
	\bibitem{OAI}
	A. M. Ozorio de Almeida and G.-L. Ingold,
	Metaplectic sheets and caustic traversals in the Weyl representation,
	J. Phys. A: Math. Theor. \textbf{47}, 105303 (2014).
	
	\bibitem{Littlejohn}
	R. G. Littlejohn,
	The Van Vleck Formula, Maslov Theory, and Phase Space Geometry,
	J. Stat. Phys. \textbf{68}, 7 (1992).
	
	\bibitem{new}
	G. M. Lando,
	Ph.D. thesis (in preparation).
	
\end{thebibliography}
\end{document}